\def\l{$\lambda$}

\def\arcmin{\hbox{$^\prime$}}
\def\arcsec{\hbox{$^{\prime\prime}$}}

\def\farcm{\hbox{$.\mkern-4mu^\prime$}}
\def\farcs{\hbox{$.\!\!^{\prime\prime}$}}  
\def\lsim{\mathrel{\hbox{\rlap{\lower.55ex \hbox {$\sim$}}\kern-.0em
\raise.4ex \hbox{$<$}}}} 
\def\gsim{\mathrel{\hbox{\rlap{\lower.55ex \hbox {$\sim$}}\kern-.0em
\raise.4ex \hbox{$>$}}}} 
\def\lya{Ly$\alpha$}
\def\grb{GRB\,021004}
\def\subsun{\mbox{$_{\odot}$}}
\def\ion#1#2{#1$\;${\small\rm\@Roman{#2}}\relax}

\documentclass{aa}
\usepackage{psfig,natbib}
\usepackage{graphicx}

\bibpunct{(}{)}{;}{a}{}{,}

\begin{document}
\title{New search strategy for high z intervening absorbers:
GRB\,021004, a pilot study\thanks{Based on observations collected at
the European Southern Observatory, Chile; proposal no.  270.A-5016}}

\subtitle{}

\author{P.~M. Vreeswijk\inst{1}
\and
P. M{\o}ller\inst{2}
\and
J.~P.~U. Fynbo\inst{3,4}
}

\offprints{pvreeswi@eso.org}

\institute{European Southern Observatory, Alonso de Cordova 3107, 
Casilla 19001, Santiago 19, Chile
\and
European Southern Observatory,
Karl-Schwarzschild-Strasse 2, D-85748, Garching bei M\"unchen, Germany
\and
Department of Physics and Astronomy, {\AA}rhus
University, Ny Munkegade, DK-8000 {\AA}rhus C, Denmark
\and
Astronomical Observatory, Copenhagen University,
Juliane Mariesvej 30, DK-2100 K{\o}benhavn K, Denmark}
\date{\today}
   
\date{Received 14 May 2003 / Accepted 6 August 2003}
   
%\authorrunning{Vreeswijk et al.}
%\titlerunning{New search strategy for high z intervening absorbers}

\authorrunning{Vreeswijk, M{\o}ller \& Fynbo}
\titlerunning{New search strategy for high z intervening absorbers:
  GRB\,021004, a pilot study}
 
\abstract{We present near-infrared narrow- and broad-band imaging of
  the field of \grb, performed with ISAAC on the UT1 of the ESO {\it
    Very Large Telescope}. The narrow-band filters were chosen to
  match prominent emission lines at the redshift of the
  absorption-line systems found against the early-time afterglow of
  \grb: [\ion{O}{iii}] at $z$=1.38 and H$\alpha$ at $z$=1.60,
  respectively.  For the $z$=1.38 system we find an emission-line
  source at an impact parameter of 16\arcsec, which is somewhat larger
  than the typical impact parameters of a sample of \ion{Mg}{ii}
  absorbers at redshifts around unity. Assuming that this tentative
  redshift-identification is correct, the star formation rate of the
  galaxy is 13$\pm$2 M\subsun\ yr$^{-1}$.  Our study reaches
  star-formation rate limits (5$\sigma$) of 5.7 M\subsun\ yr$^{-1}$ at
  $z$=1.38, and 7.7 M\subsun\ yr$^{-1}$ at $z$=1.60. These limits
  correspond to a depth of roughly 0.13 L$^*$.  Any galaxy counterpart
  of the absorbers nearer to the line of sight either has to be
  fainter than this limit or not be an emission-line source.
  } \maketitle

%
%________________________________________________________________

\section{Introduction}
\label{sec:introduction}

Identification of the objects responsible for the high redshift
intervening absorption systems in QSO spectra, such as Damped \lya\ 
(DLA) and \ion{Mg}{ii} absorbers, has been the expressed goal of a
large number of observing programmes over the past 2 decades
\citep[e.g.][]{1989ApJ...347...87S,1990A&A...236..351D,
  1997A&A...321..733L,
  1997A&A...328..499G,2001MNRAS.326..759W,2001ApJ...551...37K,
  2002ApJ...566...51C}.  Even though many such programmes have been
successful in detecting objects at (or near) the absorption redshift,
a doubt does in almost all cases persist as to whether the identified
object is indeed the absorber or if there could be another object
hidden inside the glare of the QSO itself at much smaller impact
parameter. It was only realized very late that this problem is
seriously exacerbated by the strong bias of the absorption
cross-section selection which picks out mostly objects with the
smallest impact parameters
\citep{1999MNRAS.305..849F,2001ApJ...559L...1S,2002ApJ...574...51M}.
The discovery of bright, high-redshift optical transients (OTs) of
gamma-ray bursts (GRBs) has now made it possible to completely
eliminate this last doubt.  GRB OTs, like QSOs, allow identification
of high-redshift absorbers, but the OTs, unlike QSOs, fade away
completely within a few months leaving the spectroscopically
investigated sightline open for extremely deep imaging studies.

The current status concerning the galaxy counterparts is the
following. For the high-z (z$>$2) DLA sample, the debate has been
between advocates for large, disk-forming galaxies
\citep{1986ApJS...61..249W,2001ApJ...560L..33P} and faint and small,
gas-rich dwarfs \citep{1988ApJ...329L..57T,1998ApJ...495..647H}. In
three cases DLA absorbing galaxies have been identified as faint blue
Ly$\alpha$ emitting galaxies at impact parameters ranging from 0.99 to
2.51 arcsecs (8--19 kpc), supporting the view that DLA galaxies at
z$>$2 are faint blue dwarf galaxies \citep{2002ApJ...574...51M}. In
the range from $z$=0 to $z$=1.2, Guillemin \& Bergeron (1997) reported
that galaxy counterparts of \ion{Mg}{ii} absorbers with equivalent
widths larger than 0.6\,\AA\ predominantly are spirals with
luminosities similar to present-day L$^*$ galaxies. The impact
parameters of the identified galaxies were found to correlate with the
luminosities of the galaxies, such that bright galaxies are found at
larger impact parameters. In a flux limited study, objects below the
given detection limit will fail to be identified, and because of the
correlation with impact parameter this will tend to bias all
identifications both towards brighter objects and towards larger
impact parameters. An indication that this may be the case for the
MgII studies is the clear tendency for higher redshift absorbers to be
identified as brighter galaxies with larger impact parameters (around
110 kpc at $z$=1.1) than those at lower redshifts (around 20 kpc at
$z$=0.1). The actual absorbers may therefore still be hiding at small
impact parameters below the detection limit.

GRB\,021004 is the third brightest afterglow detected so far, with
R=15.4 only 4 minutes after the burst \citep{2003Natur.422..284F}.
Spectroscopic observations \citep[e.g.][]{2002GCN..1605....1C} found a
host galaxy redshift of $z$=2.33. In addition two other systems were
identified with absorption lines of \ion{Mg}{i}, \ion{Mg}{ii} (both
systems with equivalent widhts larger than 0.6\,\AA), and \ion{Fe}{ii}
at $z$=1.38 and $z$=1.60 \citep[e.g.][]{2002A&A...396L..21M}.  In this
pilot study we present observations in narrow- and broad-band filters
to search for emission lines from the galaxy counterparts of these
absorbers.  The emission-line imaging technique that we employ allows
identification of the foreground absorber, even when it would be
located exactly along the line of sight to the GRB and its host
galaxy.  Throughout the paper we assume $\rm H_0 = 70
\,km\,s^{-1}\,Mpc^{-1}$, $\Omega_{\rm m}$ = 0.3 and $\Omega_{\lambda}$
= 0.7.

\section{Observations}

\begin{figure}
  \centering \includegraphics[width=8.5cm]{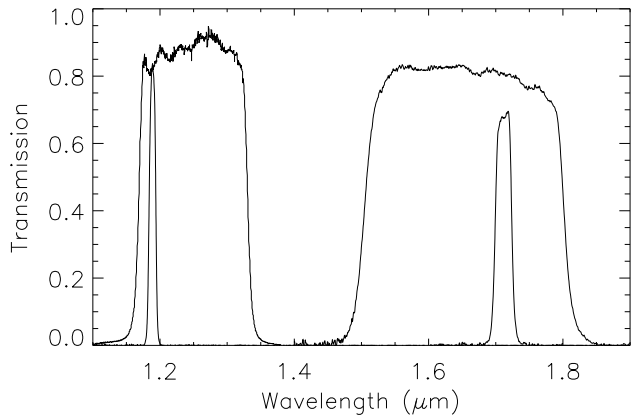}
  \caption{\small The transmission curves for the four filters used in this
  project. From left to right it is Js, NB 1.19, H, NB 1.71.
    \label{filters}}
\end{figure}

At $z$=1.38, [\ion{O}{iii}] \l 5007 is redshifted to 1.19 $\mu$m, and
at $z$=1.60, H$\alpha$ emission is shifted to 1.71 $\mu$m. These
wavelengths correspond to the central wavelengths of two ISAAC
narrow-band filters.  Imaging in these two filters and in the Js and H
band allows identification of the intervening systems if they are
star-forming galaxies.  In Fig.~\ref{filters} we show the filter
transmission curves for the four filters.

The observations were carried out in service mode with UT1/ISAAC of
ESO's {\it Very Large Telescope} on the nights of December 17/18 and
19/20, 2002, for a total effective exposure time of 51 and 105 minutes
in the narrow-band filters NB 1.19 and NB 1.71, respectively.  To
enable continuum subtraction, we also observed the field in the
broad-band filters Js and H, each for 11 minutes.  Table
\ref{tab:observations} shows the log of the observations.  The ISAAC
short-wavelength detector is a Rockwell 1024x1024 chip, with a gain of
4.5 electrons/ADU, a read noise of 11 electrons and a pixel-size of
0\farcs148.

\begin{table}[t]
  \caption[]{Log of observations}\label{tab:observations}
  $$
  \begin{array}{ccccccc}
    \hline
    \noalign{\smallskip}
    \rm filter &
    \rm obs.date &
    \rm DIT &
    \rm NDIT &
    \rm NINT &
    \rm exp.time &
    \rm seeing \\
    & 
    \rm 2002\,UT &
    \rm (sec) &
    &
    &
    \rm (sec) &
    (\arcsec) \\
    \hline
    \rm NB 1.71 & \rm Dec \,18.062 & 50 & 3 & 42   & 6300 & 0.53 \\  
    \rm NB 1.19 & \rm Dec \,20.055 & 60 & 3 & 17   & 3060 & 0.70 \\
    \rm H       & \rm Dec \,20.086 & 10 & 6 & 11   & 660  & 0.60 \\
    \rm Js      & \rm Dec \,20.098 & 30 & 2 & 11   & 660  & 0.56 \\
    \hline
  \end{array}
  $$
\end{table}

From the raw images possible ghosts of especially bright objects were
removed with the {\it ECLIPSE} \citep{2001adass..10..525D} routine
{\it ghost}. After dark subtraction, the flat-field for each image was
constructed using the {\it ECLIPSE} routine {\it twflat}.  After
flat-fielding and removal of bad pixels using the {\it fixpix} task
within {\it IRAF}\footnote{IRAF is distributed by the National Optical
  Astronomy Observatories, which are operated by the Association of
  Universities for Research in Astronomy, Inc., under cooperative
  agreement with the National Science Foundation.}, the sky was
determined for each image from the adjacent images on either side. In
the case of the narrow-band filters, we used four images on either
side, and for the broad-band filters we used as many as possible (10).
We performed two sky passes.  In the first pass all object images of a
particular Observation Block (OB) are sky subtracted. These images are
registered with a shift only (i.e. no scaling nor rotation) to the
coordinate system of the first image of the OB using about 5 bright
objects in the field, and combined. From this image we make an object
mask using the optional output image from {\it SExtractor}
\citep{1996A&AS..117..393B}, which returns the pixels above the noise
of the detected objects.  With {\it imreplace} in {\it IRAF}, these
objects were transformed into a bad pixel mask, making them larger by
a few pixels.  Then we perform another sky run, this time with all
objects masked while average-combining the adjacent sky images. Again
we register the images, and now combine them using the {\it sigclip}
rejection in {\it imcombine} within {\it IRAF}, which excludes
possible cosmic rays and bad pixels that were missed.  Finally,
residuals left from bias variations and possible horizontal jumps were
removed by subtracting the median of each row and/or column of the
image using {\it ECLIPSE} routines. This median was determined from
the final image with the objects masked. The final reduced images were
trimmed to a size of 2\arcmin\ by 2\arcmin, which is the part of the
image that was exposed in each sub-exposure (the size of the original
images is 2\farcm5 by 2\farcm5, while we set the jitter box width to
30\arcsec).

\begin{figure*}[t]
  \centering \includegraphics[width=18cm]{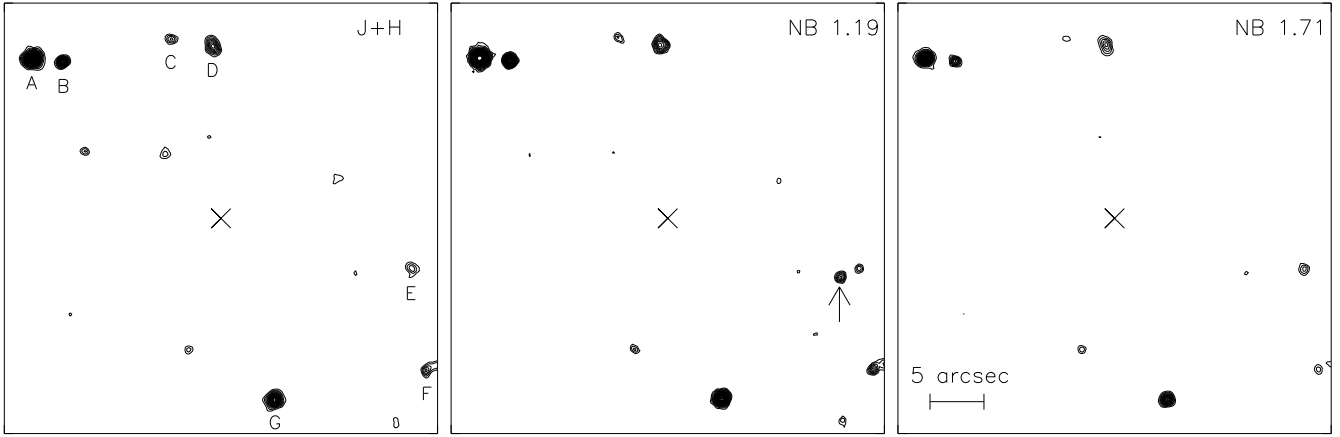}
  \caption{\small A 40$\times$40 arcsec$^2$ field centred on the position of
    \grb\ from the sum of the Js and H band images (left), the NB 1.19
    image (middle), and the NB 1.71 images (right). The only
    emission-line source closer in than 30\arcsec\ is a candidate
    [\ion{O}{iii}] emitter at an impact parameter of 16\arcsec\ seen
    near the western edge of the NB 1.19 image.
    \label{field}}
\end{figure*}

We used observations of the NICMOS standard 9123 (S427-D)
\citep{1998AJ....116.2475P} to calibrate the narrow bands and
observations of the UKIRT faint standards FS6 and FS32
\citep{2001MNRAS.325..563H} to calibrate the Js and H bands. For the
broad bands, no correction for a colour term was attempted. We
corrected for atmospheric extinction assuming an extinction of 0.06
mag per unit airmass. We estimate that the uncertainties on the
zero-points are 0.04 and 0.10 for Js and H respectively.  The
zeropoints for the narrow-band filters are determined by integrating
the spectrum of S427-D over the filter transmission curves to obtain
the flux in erg s$^{-1}$ cm$^{-1}$ outside the atmosphere.

\section{Emission-line candidates}

Inspection of the images reveals no emission in any of the filters at
the location of the early-time afterglow of \grb. To select
emission-line candidates we follow a method similar to the one
outlined in \citet{1999MNRAS.305..849F}. We let {\it SExtractor}
\citep{1996A&AS..117..393B} detect objects using a very low threshold,
i.e. requiring that 2 adjacent pixels are at least 1.5 sigma above the
sky. For all detected sources we perform aperture photometry with the
{\it phot} routine in {\it IRAF}, after which we reject all objects
with a signal-to-noise ratio less than 5. The aperture is set to a
radius of 6 pixels (0\farcs9 on the sky) for all images.  This roughly
corresponds to 1.5 times the size of the full width at half maximum
(FWHM) of the average image point-spread function (PSF). The local sky
was determined in an annulus 10 pixels wide, at a distance of 10 times
the image FWHM (about 40 pixels) from the object centre.  The
coordinates of the objects found on the narrow-band images were
transformed to the broad-band images, and at these locations aperture
photometry was performed in exactly the same manner, irrespective of
an object being present or not.

\begin{figure}[b]
  \centering \includegraphics[width=8.5cm]{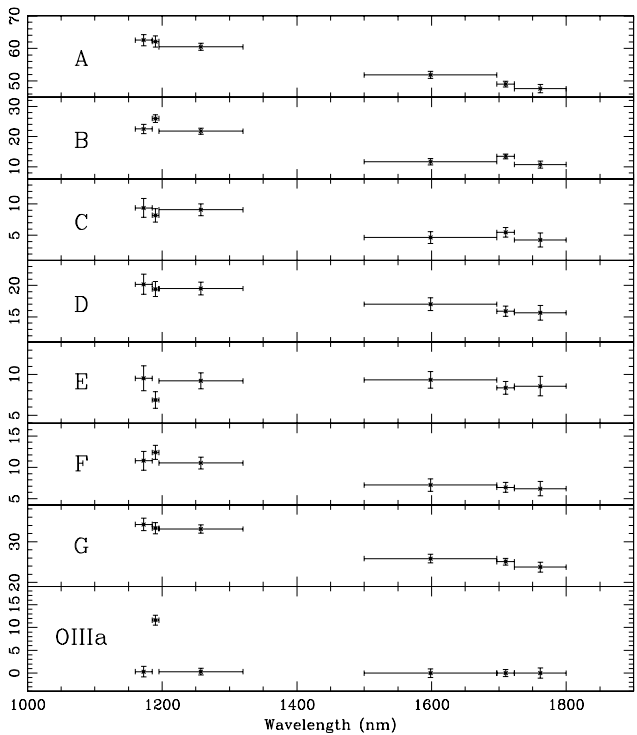}
  \caption{\small Spectral energy distributions (propotional to F$_{\lambda}$)
    of the seven objects A--F marked in Fig.~\ref{field} and for the
    candidate $z$=1.38 [\ion{O}{iii}] emitter. The candidate, named
    \ion{O}{iii}a and shown at the bottom, has strong excess flux in
    the NB 1.19 filter.
    \label{spec}}
\end{figure}

In both narrow-band images we detect a number of emission-line
candidates (6 in NB 1.19 and 4 in NB 1.71 above 5$\sigma$), but almost
all at very large impact parameters ($>$30\arcsec).  Only one
candidate in the NB 1.19 image, i.e.  [\ion{O}{iii}] at $z$=1.38, has
a small enough impact parameter to be a possible galaxy counterpart of
the absorber consistent with the sample identified by
\citet{1997A&A...328..499G}.  It is located at 16\arcsec\ from the GRB
line-of-sight and can be seen at the western edge of the section of
the field shown in Fig.~\ref{field}. In Fig.~\ref{spec} we show the
spectral energy distributions of the seven continuum objects named
A--F in Fig.~\ref{field} and of the [\ion{O}{iii}] emission-line
candidate named \ion{O}{iii}a. Although the emission-line nature of
this object is clear, follow-up spectroscopy is needed to verify the
redshift.

The NB 1.19 emission-line flux of \ion{O}{iii}a is 8.1 $\times$
10$^{-17}$ erg s$^{-1}$ cm$^{-2}$, while its continuum 3$\sigma$ upper
limit, estimated from the Js and H upper limits, is 2.1 $\times$
10$^{-17}$ erg s$^{-1}$ cm$^{-2}$ in the NB1.19 filter.  Under the
assumption that the emission line is indeed [\ion{O}{iii}], the
candidate is at a luminosity distance of 3.0 $\times$ 10$^{28}$ cm and
has an [\ion{O}{iii}] luminosity of L$_{\rm [\ion{O}{iii}]}$ = 7
$\times$ 10$^{41}$ erg s$^{-1}$. Using the empirical flux ratio
[\ion{O}{iii}]/H$\alpha$ of 1/2 (based on Table~5 of
\citet{1997A&A...328..499G}), we obtain an estimated H$\alpha$
luminosity of 1.4 $\times$ 10$^{42}$ erg s$^{-1}$. This corresponds to
a star-formation rate (SFR) of 13$\pm 2$ M\subsun\ yr$^{-1}$
\citep{1998ARA&A..36..189K}.  The error given here is the measurement
error; an additional error of order 30\% originates from the
uncertainty in the H$\alpha$ luminosity to SFR conversion factor
\citep{1998ARA&A..36..189K}.

Our observations are sensitive to a limiting luminosity (5$\sigma$
with an aperture radius of 6 pixels, and assuming the continuum
contribution in the narrow-band filters is negligible compared to the
emission line) of L$_{\rm [\ion{O}{iii}]}$ = 3.6 $\times$ 10$^{41}$
erg s$^{-1}$ at $z$=1.38, and L$_{\rm H\alpha}$ = 9.7 $\times$
10$^{41}$ erg s$^{-1}$ at $z$=1.60. The H$\alpha$ luminosity is about
0.13 L$^*$ relative to the Schechter parametrisation of the H$\alpha$
luminosity function of galaxies at similar redshifts of
\citet{1999ApJ...519L..47Y}.  These luminosity limits correspond to
star-formation rates of SFR$_{z=1.38}$=5.7 M\subsun\ yr$^{-1}$ and
SFR$_{z=1.60}$=7.7 M\subsun\ yr$^{-1}$ \citep{1998ARA&A..36..189K}.

%Any galaxy counterpart of the absorbers nearer to the line of sight
%than \ion{O}{iii}a either has to be fainter than these limits or not
%be an emission-line source.

\section{Discussion}

The 16\arcsec\ impact parameter of the NB 1.19 candidate corresponds
to 135 kpc at $z$=1.38.  The sample of quasar absorption-line systems
identified with galaxies by \citet{1997A&A...328..499G} were found to
have separations of 1.9--23\arcsec\ which, in the cosmology that we
assume, correspond to 12--127 kpc. Our candidate is therefore
consistent with, but in the upper range of, the impact parameters of
the \ion{Mg}{ii} systems previously studied.  However, we find 9
emission-line objects down to a limiting SFR of 7.7 M\subsun\ 
yr$^{-1}$ in two fields each of angular size 2\arcmin$\times$2\arcmin,
so there is a priori 50\% chance that one of those objects would be
found within an impact parameter of 16\arcsec\ in one of the fields.

In conclusion, this pilot study is consistent with the result of
\citet{1997A&A...328..499G} that the typical \ion{Mg}{ii} absorber is
an L$^*$ galaxy at an impact parameter of 100 kpc. In our case the
candidate absorber is less luminous, but in the upper end of the
impact parameter distribution.  In their intermediate redshift
(0.7--1.2) sample, Guillemin \& Bergeron identified 10 of 16 absorbers
but failed to find counterparts to the remaining 6.  We find one but
fail to identify a candidate in the other field, while our
star-formation limit (see above) compared to theirs
(SFR$_{z=0.7-1.3}$=0.7--43 M\subsun\ yr$^{-1}$) is such that we
would have expected to detect half of the objects they identified.

The question therefore remains open whether these large galaxies are
really the absorbers, or if they are merely the brightest neighbour to
dwarf galaxy absorbers. For DLA absorbing galaxies at higher redshifts
impact parameters are found to be of order 10 kpc rather than 100 kpc
\citep{2002ApJ...574...51M}, and the SFRs of order 5--7 M\subsun\ 
yr$^{-1}$. This is just below the 5$\sigma$ limit of our current data.
A future larger study of GRB intervening \ion{Mg}{ii} and DLA
absorbers should aim at reaching at least a factor of three deeper (in
SFR) in order to answer the question if we are still missing a
population of faint absorbers at low impact parameters, and if
\ion{Mg}{ii} absorbers are in fact related to dwarf galaxies of the
DLA type.

\begin{acknowledgements}
  We are grateful to the ESO staff in Garching and Paranal, in
  particular to Mario van den Ancker and Rachel Johnson, for allowing
  prompt execution of our proposal.  PMV is also thankful to John
  Willis, Chris Lidman and Sara Ellison for very helpful discussions.
  JPUF gratefully acknowledges support from the Carlsberg Foundation.
\end{acknowledgements}

\bibliographystyle{aa}
\bibliography{references}

\end{document}